\documentclass[pre,twocolumn,twoside,superscriptaddress,floatfix]{revtex4-1}

\usepackage{currfile}
\usepackage{color}

\usepackage{graphicx,epsfig,verbatim,enumerate}
\usepackage{amssymb,amsmath}
\usepackage{ifthen}
\usepackage{hyperref}

\usepackage[sc]{mathpazo}
\usepackage[T1]{fontenc}
\usepackage{microtype}
\usepackage{textcomp}

\usepackage{longtable}

\usepackage{mathtools}

\newboolean{twocolswitch}

\newcommand{\sindex}[1]{}
\newcommand{\nindex}[1]{}

\newcommand{\www}[1]{\url{#1}}

\usepackage{lettrine}

\graphicspath{{figures/}}

\setboolean{twocolswitch}{true}

\begin{document}

\title{\protect
Connecting every bit of knowledge: \\
The structure of Wikipedia's First Link Network
}

\author{
\firstname{Mark}
\surname{Ibrahim}
}
\email{mark.s.ibrahim@uvm.edu}

\affiliation{Department of Mathematics \& Statistics, 
    Computational Story Lab,  \\
    Vermont Complex Systems Center,
    Vermont Advanced Computing Core,
    The University of Vermont, Burlington, VT 05401.}

\author{
\firstname{Christopher}
\surname{M. Danforth}
}
\email{chris.danforth@uvm.edu}

\affiliation{Department of Mathematics \& Statistics, 
    Computational Story Lab,  \\
    Vermont Complex Systems Center,
    Vermont Advanced Computing Core,
    The University of Vermont, Burlington, VT 05401.}

\author{
\firstname{Peter}
\surname{Sheridan Dodds}
}

\email{peter.dodds@uvm.edu}

\affiliation{Department of Mathematics \& Statistics, 
    Computational Story Lab,  \\
    Vermont Complex Systems Center,
    Vermont Advanced Computing Core,
    The University of Vermont, Burlington, VT 05401.}

\date{\today}

\begin{abstract}
  \protect
  Apples, porcupines, and the most obscure Bob Dylan song---is every topic a few clicks from Philosophy? 
Within Wikipedia, the surprising answer is yes: nearly all 
paths lead to Philosophy.
Wikipedia is the largest, most meticulously indexed collection of human knowledge ever amassed. 
More than information about a topic, Wikipedia is a web of naturally emerging relationships.  
By following the first link in each article, we algorithmically construct a directed network of 
all 4.7 million articles: Wikipedia's First Link Network.
Here, we study the English edition of Wikipedia's First Link Network for insight into how the many 
articles on inventions, places, people, objects, and events are related and organized.  

By traversing every path, we measure the accumulation of first links, path lengths,
groups of path-connected articles, and cycles.
We also develop a new method, traversal funnels, to measure the influence each article exerts in shaping the network. 
Traversal funnels provides a new measure of influence for directed networks without spill-over into cycles, in contrast to traditional network centrality measures.
Within Wikipedia's First Link Network, we find scale-free distributions describe path length, 
accumulation, and influence. Far from dispersed, first links disproportionately accumulate 
at a few articles---flowing from specific to general and culminating around fundamental notions such as
Community, State, and Science. 
Philosophy directs more paths than any other article by two orders of magnitude. 
We also observe a gravitation towards topical articles such as 
Health Care and Fossil Fuel. 
These findings enrich our view of the connections and structure of
Wikipedia's ever growing store of knowledge.

\end{abstract}


\maketitle

\section{Introduction}

Wikipedia is a towering achievement of the Internet age. 
At no point in history has a larger or more meticulously indexed collection of human knowledge 
existed.
Wikipedia contains 37 million articles in 283 languages, 
with coverage spanning everything from little known ancient battles to the latest pharmaceutical drugs 
\cite{clauson2008scope, stats}.
Demonstrating its relevance to modern inquiry,
Wikipedia is the sixth most visited
site in the world, surpassing 18 billion page views and 10 million edits in January, 2013 alone.
\cite{wiki_edits, wiki_views}.

Wikipedia has naturally become the object of many studies. 
Researchers have examined the cultural dynamics among editors
\cite{iba2010analyzing, samoilenko2016linguistic},
the accuracy of the content relative to traditional encyclopedias
\cite{holman2008comparison, giles2005internet},
the topics covered 
\cite{halavais2008analysis},
and bias against portions of the population
\cite{hill2013wikipedia}.
Wikipedia's content has also proven to be a powerful tool. 
Researchers have used Wikipedia to identify missing dictionary entries
\cite{williams2015identifying},
cluster short text
\cite{banerjee2007clustering},
compute semantic relatedness
\cite{gabrilovich2007computing},
and disambiguate meaning \cite{cucerzan2007large}.

While these many studies have dissected and fruitfully applied Wikipedia's content,
here we examine the connections among the many articles.
A hyperlink from one Wikipedia article to another
naturally indicates a relationship between the two articles
\cite{kamps2009wikipedia}.
The notion that hyperlinks convey information about the content of a
page has proved enormously successful in multiple domains from search engine algorithms 
such as PageRank 
\cite{page1999pagerank} 
to topic classification
\cite{chakrabarti2001integrating}.
Here, we treat a hyperlink as a mechanism connecting two topics.

The authors of a Wikipedia article choose where and whether to include a 
reference to another Wikipedia article
in the HTML markup.
For example, as of November 2014, the authors of the 
\href{https://en.wikipedia.org/wiki/Train}{``Train''}
article
had collectively chosen
``Amtrak's Acela Express,'' ``steam,'' and ``head-end power'' 
among others as relevant articles to reference in describing ``Train''
\cite{wiki_train}
.

\begin{figure*}[tp!]
  \includegraphics[width=\textwidth]{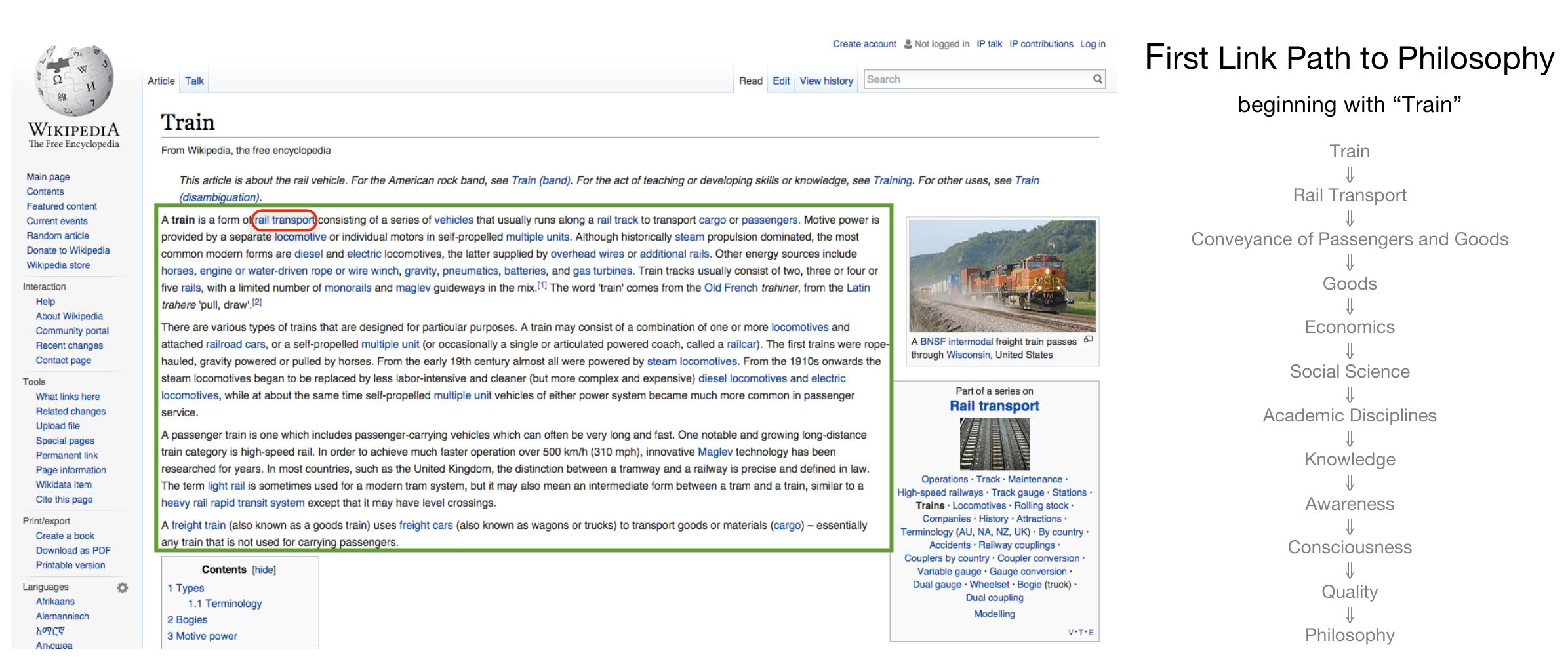}  
  \caption{
    \textbf{First Link Path For ``Train.''}
    We follow the first link  to another Wikipedia article
    in the main body of the article---the area
    inside the green rectangle, which excludes 
    side bar elements, the navigation bar and title; the first link is circled in red. 
    In this example, the first link to another Wikipedia article is ``Rail Transport.'' 
    We can again select the first link on the ``Rail Transport''
    article, repeating the process to 
    form a path of first links. After 11 links, we arrive at ``Philosophy.''
  }
  \label{fig:Train First Links}
\end{figure*}

By focusing our attention on the main body of an article---excluding
elements in side bars and headings---we 
attempt to systematically capture the core description of a topic.
Within the body text, the first link marks the earliest moment in a topic's introduction where the 
authors choose to directly reference another article. 
While many links reference relevant details, 
the first link is an association within a topic's initial 
description. While 
``Amtrak's Acela Express,'' ``steam,'' and ``head-end power'' 
are links detailing particulars, the first link, ``rail transport,''
is the topic the authors associate with ``Train'' in the introduction
(Fig.~\ref{fig:Train First Links}).
``Banana'' has a first link to ``fruit,'' ``Bob Dylan'' has a first link to 
``Blowing in the Wind,'' and ``Physics'' has a first link to ``natural science.''
Collectively, first links provide a pragmatic and interpretable 
means to connect each article to another.

By following the first hyperlink contained in all articles of the English edition of
Wikipedia as of November, 2014, we form a directed network:
{\it Wikipedia's First Link Network} (FLN).
Inspired by the claim that the majority of first links lead to
``Philosophy''---popularized by an 
\href{https://xkcd.com/903/}{xkcd}
comic and subsequently
discussed in blog posts 
\cite{xkcd, mat_blog, Ilmari_first_links, 
bob_west, xefer2011philosophy}
---
we holistically study how the FLN  yields a flow of connections.
For methodological details, see Sec.~\ref{Appendix}. 

The FLN is a wealth of relations among inventions, places,
{figures, objects, and events across space and time.
``Train'' for example, links to a parent node, ``Rail Transport,''
while many child nodes such as ``Steel'' and ``Horsepower'' link to ``Train.''
As shown in Fig.~\ref{fig:Train First Links}, the path starting at ``Train'' 
contains ``Goods,'' ``Economics,'' ``Social Science,''
leading ultimately to ``Philosophy''.
Unlike previous taxonomies created by individuals
\cite{Bolton2007, descartes, aristotle, hist_thesaurus},
the relations in the FLN emerge without a centralized effort 
as the aggregate of each article's authors choice of first link.

Our goal is to study the structure of the FLN for insight into how the information on Wikipedia is organized and related.
Informed by river network metrics,
we consider the FLN as a flow. We quantify
the accumulation of first links around articles 
and develop a new method, \textit{traversal funnels}, to measure the influence an article exerts in shaping the FLN.
Together with cycles, in-degree, depth, and the content of the articles, 
we build our analysis of the relations among the ideas in Wikipedia.

\section{Traversing the First Link Network}
\label{Traversal Algorithm}

An essential feature of a directed network's structure is the degree distribution 
\cite{newman2003structure}. 
The degree distribution has been used to study many phenomena from disease outbreak 
\cite{eubank2004modelling} 
to the dynamics of social networks 
\cite{newman2002random}.
The in-degree distribution in the FLN describes how many first links point to a 
particular article. 
Articles with zero in-degree have no references---they are outer leaves in the FLN 
or are disconnected entirely. 

\begin{figure*}[tp!]
  \includegraphics[width=\textwidth]{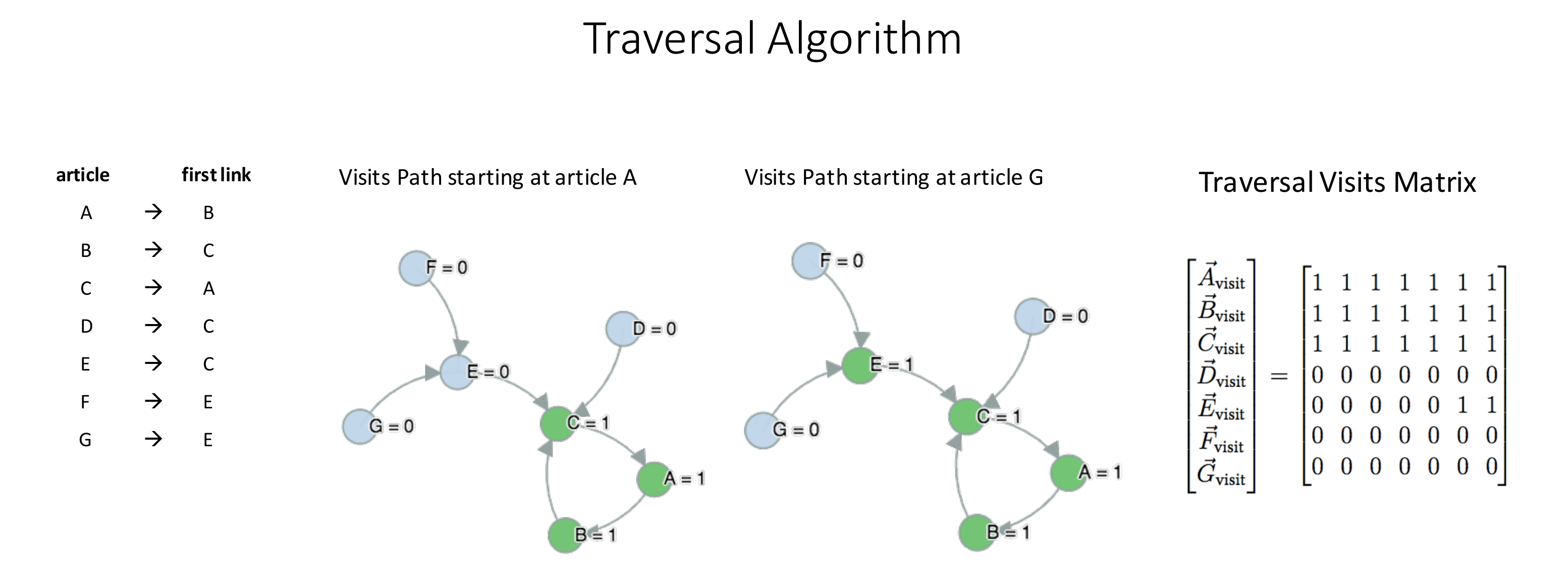}  
  \caption{
    \textbf{Traversal Visit Algorithm on a sample network.}
     The traversal visit vectors are an adjacency matrix for the paths through the network: 
     the first column indicates the path formed starting with article A. 
     The number of traversal visits for article A is then the number of paths containing A 
     or the sum of the first row in our matrix:
     $\sum_{i=1}^7 A_{\text{visit, i}} = 7$.
  }
  \label{fig:Traversal Visits}
\end{figure*}
Starting at each article, we construct a path through the FLN, 
to map the flow of connections among articles.
The method is order agnostic with respect to which articles are selected first. As long as each article is selected eventually, the resulting metrics are equivalent.
Previous studies have used flow to characterize the structure of river networks
\cite{horton1945erosional, dodds1999unified} and describe the organization of 
blood networks, food systems, and transportation networks
\cite{garlaschelli2003universal}. 
In the same vein, we use flow to measure accumulation and develop a new method for isolating influence in a directed network with cycles.

\subsection{Traversal Visits}
\label{Traversal Visits}

The first metric we develop quantifies the accumulation of first links.
The algorithm begins by selecting an article, then traversing the path formed
by following the first links. Each time a first link references an article, we increment a count
associated with the article. 
We continue until a page is revisited or is invalid---defined here to mean pointing to a page outside of Wikipedia.
We select a second article and repeat the process until we have 
constructed a path for each article in the network. 
We define the number of {\it traversal visits} of an article 
to be the number of references flowing to the ideas in the article---equivalent
to drainage basin area in geomorphology. 

We can characterize the paths in the FLN as a matrix with each column corresponding to a path. In our sample network 
(Fig.~\ref{fig:Traversal Visits}), the path starting at article A is the 
first column in the traversal visits matrix. 
An entry of $1$ indicates the path contains a given article and 
$0$ indicates the path does not.
To compute the number of traversal visits for an article, we sum the corresponding row in our matrix.
The traversal visits matrix for 
our Wikipedia dataset consists of 121 million entries encoding each path out of the more than 
googol possible paths ($\simeq4.7\times10^6\times2^{4.7\times10^6}$) through the FLN .

By measuring the number of first links between
two articles, we obtain an additional piece of information we call 
{\it path length}, computed by summing 
columns in the traversal visits matrix. 
Path length describes how closely related topics are.
Although ``Train'' is related to ``Economics'' for example, there are several articles
bridging the connection:
``Train'' is more specifically related to transportation, whose object
is often goods. Goods are one of the fundamental objects of study in Economics.
Described in links, this relationship is captured by a relatedness of $4$ 
first links ultimately connecting ``Train'' to ``Economics.'' 

One possible path through the FLN is a {\it cycle} or a group of articles 
linking to one another inside a loop. In our sample network 
(Fig.~\ref{fig:Traversal Visits}) 
a 3-cycle exists among nodes A, B, and C. 
We can readily identify
the types of cycle structures within the FLN
and rank each by 
the number of references directed towards a cycle. 
We can also identify and rank {\it groups of 
path-connected articles}, not necessarily forming a perfect cycle.

\begin{figure}[tp!]
  \includegraphics[width=\columnwidth]{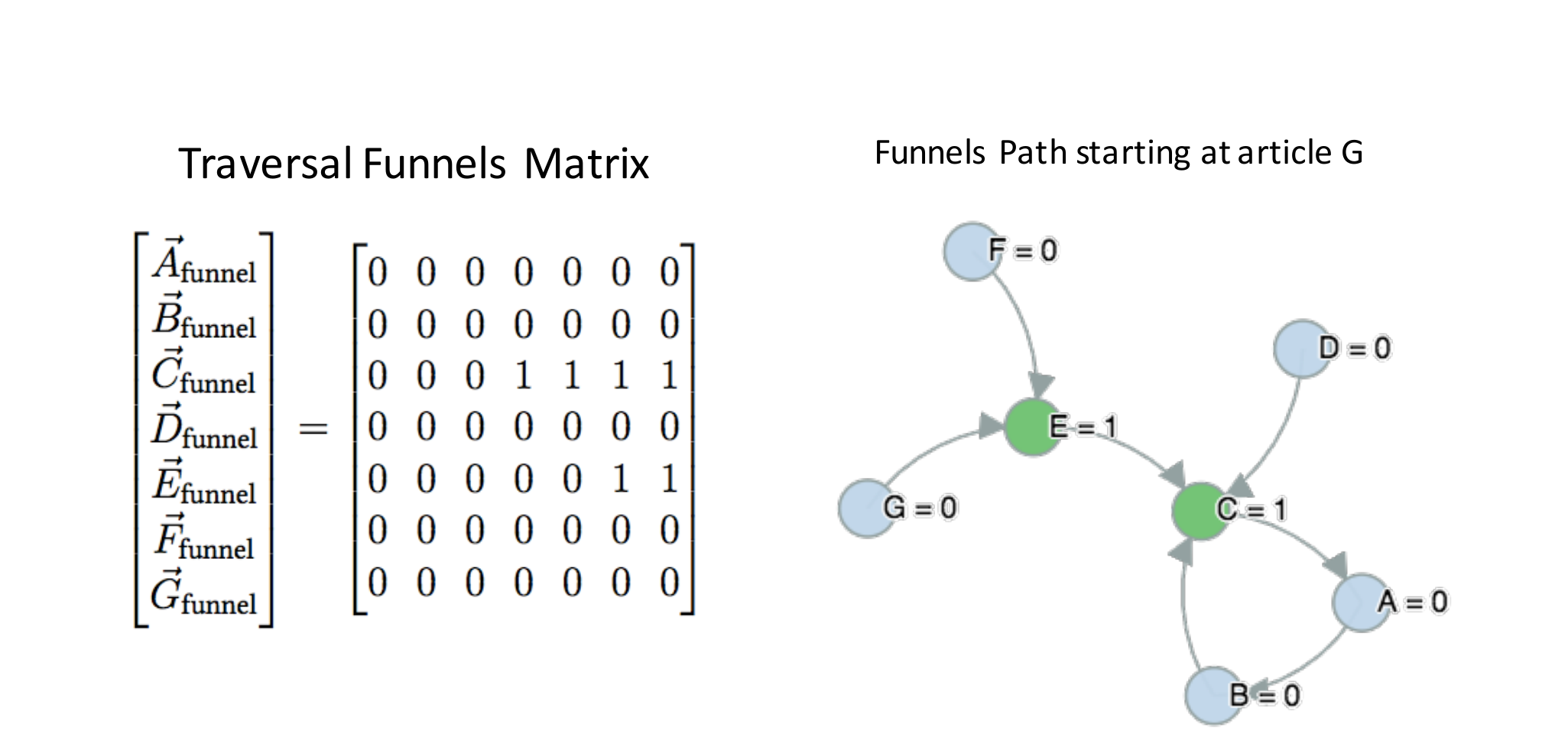}
  \caption{
    \textbf{Traversal Funnel Algorithm on a simple network.}
  The algorithm for traversal funnels is identical to the previous algorithm for traversal visits with one alteration: the path ends at the start of a cycle to distinguish articles directing a path into a cycle from articles that simply happen to be in a highly traversed path. We can construct similar vectors by considering each path through the network, measuring traversal funnels for a particular article as the sum of the entries in its corresponding row. For example
  the number of traversal funnels for article $E$ is 
  $\sum_{i=1}^7 E_{\text{funnel, i}} = 2$.}
  \label{fig:Traversal Funnels}

\end{figure}
\subsection{Traversal Funnels}
\label{Traversal Funnels}

\begin{figure}[tp!]
  \includegraphics[width=\columnwidth]{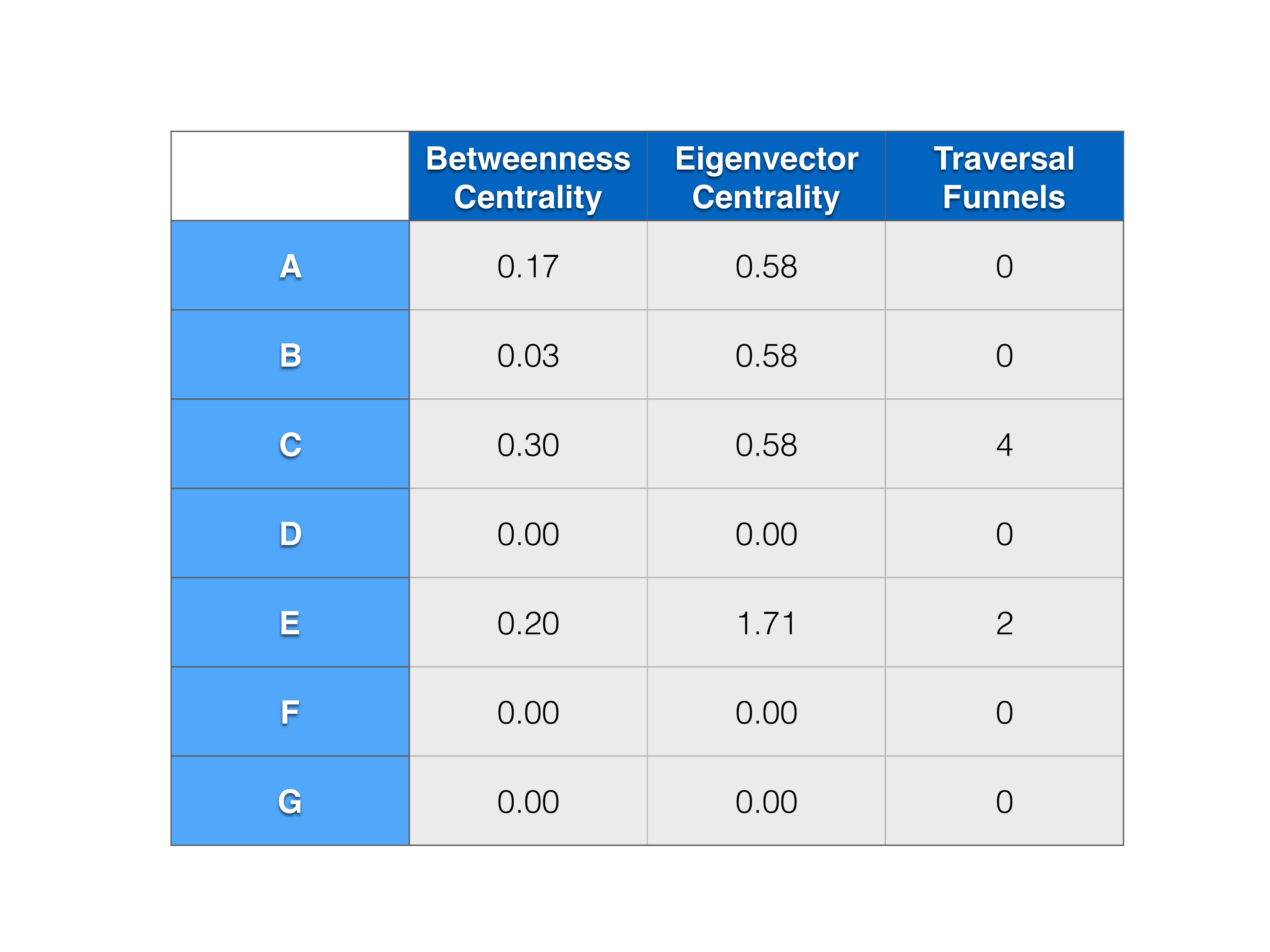}
  \caption{
    \textbf{Centrality Measures}
    The table above compares standard centrality measures on the sample network against traversal funnels. While betweenness centrality and eigenvector centrality both reward node A for its place in the cycle, traversal funnels only rewards node C for directing paths into the cycle. }
  \label{fig:Centrality Measures}

\end{figure}

While traversal visits measure accumulation, each article's first link also 
influences the shape of the FLN. 
At a point of accumulation, a single article's first link 
can exert great influence over the shape of the FLN by directing many
references on a particular path. To distinguish between an article 
that simply happened to fall within a cycle from an article funneling 
many first links, we develop a second metric called {\it traversal funnels}.
To measure traversal funnels, we traverse the FLN in the same manner as we 
did for traversal visits, but end a path once we enter a cycle.
We are then able to distinguish between an article related to many other ideas
only by virtue of its place in a cycle, from an article exerting influence over where the first links flow. 
In the sample network 
(see Fig.~\ref{fig:Traversal Funnels}), article C 
directs the flow of links towards the 3-cycle, while articles A and B are 
recipients of the flow---without exerting direct influence themselves. 

The algorithm increments the traversal funnels count along each path, up to the start of a cycle. The resulting count is each node's number of traversal funnels: the number of paths the node directs up to the start of a cycle. Below is the pseudo-code. 

\begin{verbatim}
Traversal Funnels Algorithm

// 1. Determine Cycles 
for node in Graph
    next_node = node.link
    visited = {node, next_node}
    // advance to potential cycle
    while next_node not in visited:
        next_node = next_node.link
        visited.add(next_node)
    // mark cycle
    if (node == next_node):
        node.in_cycle = True

// 2. Compute Traversal Funnels
for node in Graph:
    while node.in_cycle == False:
        node.funnels += 1
        node = node.link
\end{verbatim}

The algorithm's average runtime is proportional to the product of the number of nodes, $n$, and the average path length, $l$. If the average path length is fixed (independent of $n$), the runtime is $O(n)$ with $O(n + l)$ auxiliary memory for storing cycles and traversing paths. In the worst case where $l$ is proportional to $n$, the runtime is $O(n^2)$ with $O(n)$ auxiliary memory for storing cycles and traversing paths. In implementation, each iteration step in the algorithm can be parallelized since the algorithm is order agnostic, as long as each node is eventually selected.

As a measure of influence, traversal funnels differs from traditional centrality measures such as degree centrality, betweenness centrality, and eigenvector centrality. 
Degree centrality characterizes how many articles have a first link to a particular topic, restricting influence to a single link. 
Betweenness centrality and eigenvalue centrality capture flow within the network more than a single link away. 
Unlike traversal funnels however, betweenness centrality and eigenvalue centrality reward nodes that simply happen to be in a cycle. 
We compute betweenness centrality and eigenvalue centrality on the sample network using \href{https://networkx.github.io/}{NetworkX}, a Python package for modelling complex networks (see the \href{http://compstorylab.org/share/papers/ibrahim2016a/index.html}{online appendix} for details). 
In our sample network (see Fig.~\ref{fig:Centrality Measures}), node A has betweenness centrality of 0.17 and eigenvalue centrality of 0.58, although node A is simply the recipient of paths directed by node C.
Eigenvector centrality even assigns the same centrality to nodes A, B, and C. 

In contrast, traversal funnels defines a path without a predetermined end node, following the progression of links to its culmination at the start of a cycle. Furthermore, traversal funnels provides an interpretable metric: a node's traversal funnels is the number of paths the node directs to the start of a cycle. 
Consequently, traversal funnels allow us to isolate the paths directed by node C, of which there are 4, without any spill over effect on other nodes in its cycle: A and B have zero traversal funnels. 

In addition, traversal funnels extends to directed networks with more than one outward edge. In such a network, there may be multiple paths associated with a single starting node. 
The same interpretation for traversal funnels as the number of paths a node directs to a cycle still holds. 
In contrast to betweenness centrality, traversal funnels would not favor the shortest path over any other. Traversal funnels equally weighs all paths up to a cycle. 

By considering the FLN not only as collection of directly linked pairs of articles, but
as a flow using traversal funnels, with accumulated references, cycles, and path-connected groups, we explore how the many articles in Wikipedia are organized and related.

\section{Results}
\subsection{Degree Distribution}

\begin{figure}[tp!]
  \includegraphics[width=\columnwidth]{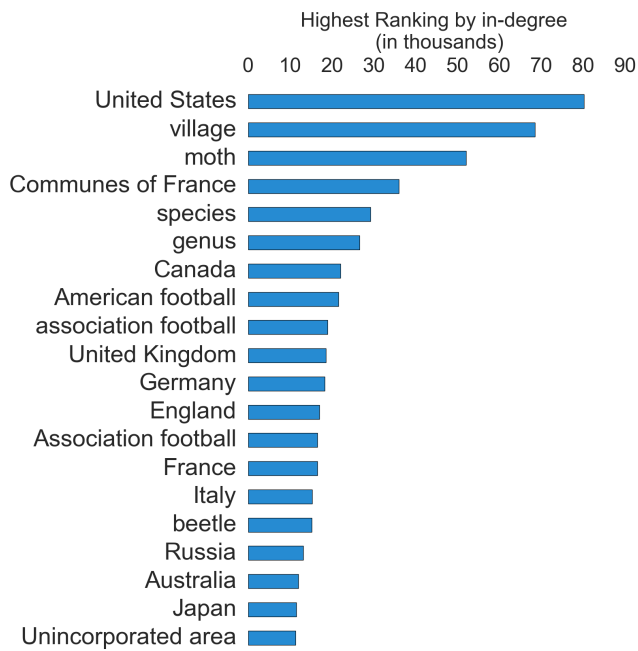}
  \caption{
    \textbf{Highest Ranking Articles by in-degree.}
    We rank each article by the number of direct first links to the article (in-degree). The highest-ranking articles tend to represent geographical and biological abstractions. A full online appendix of the results and data is available \href{http://compstorylab.org/share/papers/ibrahim2016a/index.html}{here}.}
  \label{fig:indegree list}
\end{figure}
We rank all 11 million articles in the FLN by in-degree to find 
the ``United States'' with 80,249 direct first links as the most referenced
Wikipedia article 
(Fig.~\ref{fig:indegree list}). 
Other high-ranking articles
include foundational abstract concepts such as ``village,'' ``species,''; 
sports associations such as ``American Football,'' ``Association Football''; 
and developed nations such as ``France,'' ``Japan,'' ``Russia,'' ``Australia,'' and 
the ``Netherlands.'' These high ranking articles are useful abstractions: nations
describe a collection of individuals with a common culture, language, or 
geographical proximity; sports teams describe an ever changing collection of 
players often associated with a cultural identity or a geographical 
region. 
Since abstractions such as nations and teams are inherently comprised
of many parts, authors often reference the abstraction when describing a part.

``Philosophy'' and other philosophical concepts
are not among the highest-ranking articles by in-degree.
``Philosophy'' has an in-degree of only 581, with direct first links from articles about Philosophers and areas of Philosophy: ``Existentialism and Humanism,'' ``Predeterminism,'' ''Synoptic Philosophy,'' ``Qualia,'' ``Dorothy Emmet,'' and ``Christopher W. Morris.''
So while a great many FLN paths flow to 
``Philosophy'' (see traversal visits discussion below), 
the accumulation is not the 
result of many articles directly referencing ``Philosophy.'' 
Instead, first links flow towards ``Philosophy'' as the 
ultimate anchor, by generalizing from specific to broad.

The FLN's in-degree exhibits a decaying power law distribution where a few articles 
receive most direct first references, while most articles receive few or none.
The average in-degree for all 11 million articles is 3.6 direct first links with a standard deviation of 89.5.
Less than $1\%$ of articles have more than 100 direct first links and $75\%$ of articles
have fewer than 9. 
We plot the in-degree distribution against each article's rank with a log-log (base 10) and fit the data to a power law 
distribution using the Maximum Likelihood Estimator with xmin as the in-degree minimizing Kolmogorov-Smirnov distance
(see Fig.~\ref{fig:degree distribution}) 
\cite{clauset2009power, alstott2014powerlaw}
.
A power law distribution obeys $Y = r^{-\alpha}$ where $r$ is the rank and $\alpha$ is Zipf's rank
exponent
\cite{zipf1949human}
. We can also examine the size distribution by measuring the number of articles, size $S$, above
a particular in-degree, which is distributed according to: $S^{-\gamma}$ ($\gamma$ is the size exponent). 
The two exponents obey the relationship: $\alpha = \frac{1}{\gamma -1}$.
For in-degree, we find xmin to be 5 links with a corresponding power law rank exponent $\alpha \simeq 2.44$ and a size exponent $\gamma \simeq 1.41$. 
We also compare the fit against alternative distributions using the log likelikhood ratio of a Kolmogorov-Smirnov Test 
(see Sec.~\ref{power_law_comp}).

\begin{figure}[tp!]
  \includegraphics[width=\columnwidth]{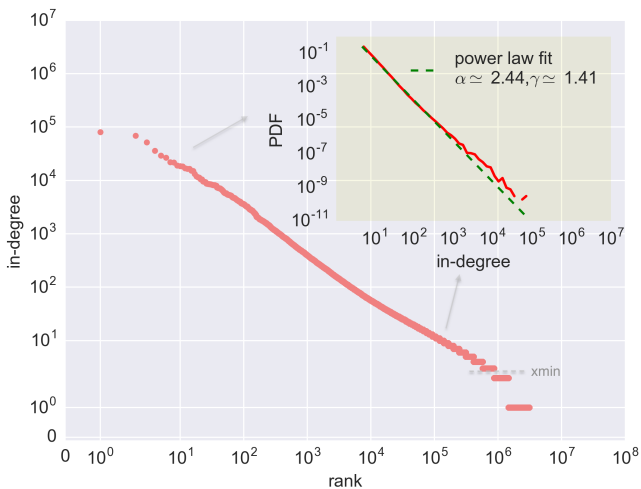}
  \caption{
    \textbf{FLN Degree Distribution.}
    We plot the in-degree distribution against rank for all articles with an in-degree of at least 1 on a $\log_{10}-\log_{10}$. We then fit the data to a power law 
    distribution using the Maximum Likelihood Estimator with xmin as the in-degree minimizing Kolmogorov-Smirnov distance. We find xmin to be 5 links with a corresponding power law rank exponent $\alpha \simeq 2.44$ and a size exponent $\gamma \simeq 1.41$.
}
  \label{fig:degree distribution}
\end{figure}

\subsection{Depth of the FLN}

\begin{figure}[tp!]
  \includegraphics[width=\columnwidth]{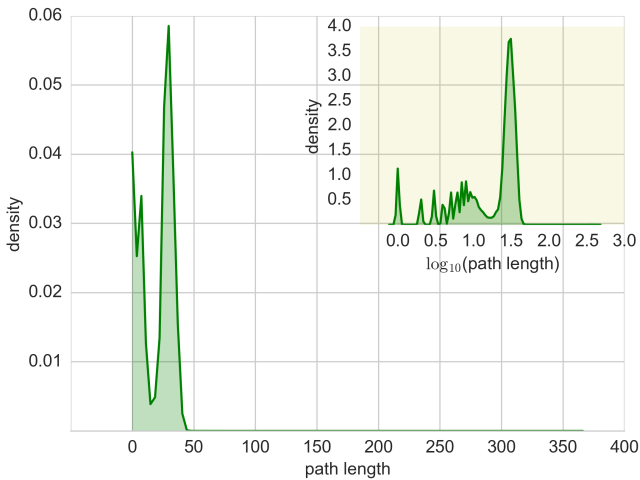}
  \caption{
    \textbf{Path Length Distribution.}
The network depth appears to have a bimodal distribution with a median of 29 first links in a path.
The 365 cycle for ``Orthodox'' liturgics is the outlier to the right, with historical articles 
describing nations such as ``Scotland'' and ``UK'' also lying outside the third quartile.
More than $75\%$ of articles have path lengths between 
$0$ and $50$ links.}
  \label{fig:Path Length Distribution}
\end{figure}
How many links does a connection among articles span? 
We find the longest path length is 365,
corresponding to the yearly calendar of Orthodox Liturgics.
Each day's Liturgics links to the next day's. On the last calendar day, the last article simply links back to January 1, forming a 365-cycle 
(see discussion of cycles ~\ref{cycles}).
We also find similarly lengthy paths following the evolution of a place or topic through time: 
``1953 in Scotland'' or ``1560s Architecture,'' with articles sequentially proceeding by year, decade or era.
In general, the longest paths connect temporally organized ideas.

Of the 11 million articles, 5.5 million had an invalid link or linked back to the same article, yielding a path length of zero. 
This roughly corresponds to the official number of articles on Wikipedia: 
$~4.7$ million as of November 2014---approximately half of the 11 million 
articles in the XML dump are redirects or disambiguations, not full articles.
The most common path length is 29, with $75\%$ of path lengths ranging between 7--30 articles.
As a distribution, more than $75\%$ of articles have a path length below 
$50$ first links 
while a few temporally organized paths exceed 50 links 
(see Fig.~\ref{fig:Path Length Distribution}). 

\begin{figure}[tp!]
  \includegraphics[width=\columnwidth]{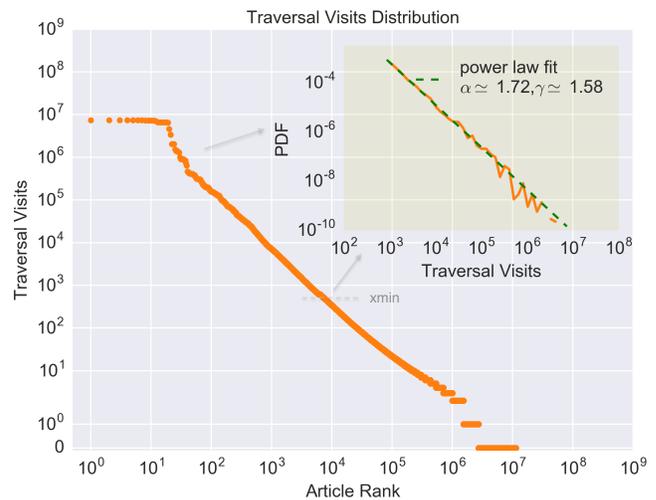} 
  \caption{
    \textbf{Distribution of Traversal Visits.}
    We plot the number of traversal visits for each article by rank on a $\log_{10}-\log_{10}$. We then fit the data to a power law 
    distribution using the Maximum Likelihood Estimator with xmin as the number of traversal visits minimizing Kolmogorov-Smirnov distance. We find xmin to be 866 with a corresponding power law rank exponent $\alpha \simeq 1.72$ and a size exponent $\gamma \simeq 1.58$.
The horizontal flattening around the highest
ranking articles is a result of the cyclic structure (see discussion of Cycles ~\ref{cycles}).
}
  \label{fig:Distribution of Visits}
\end{figure}

\begin{figure}[tp!]
  \includegraphics[width=\columnwidth]{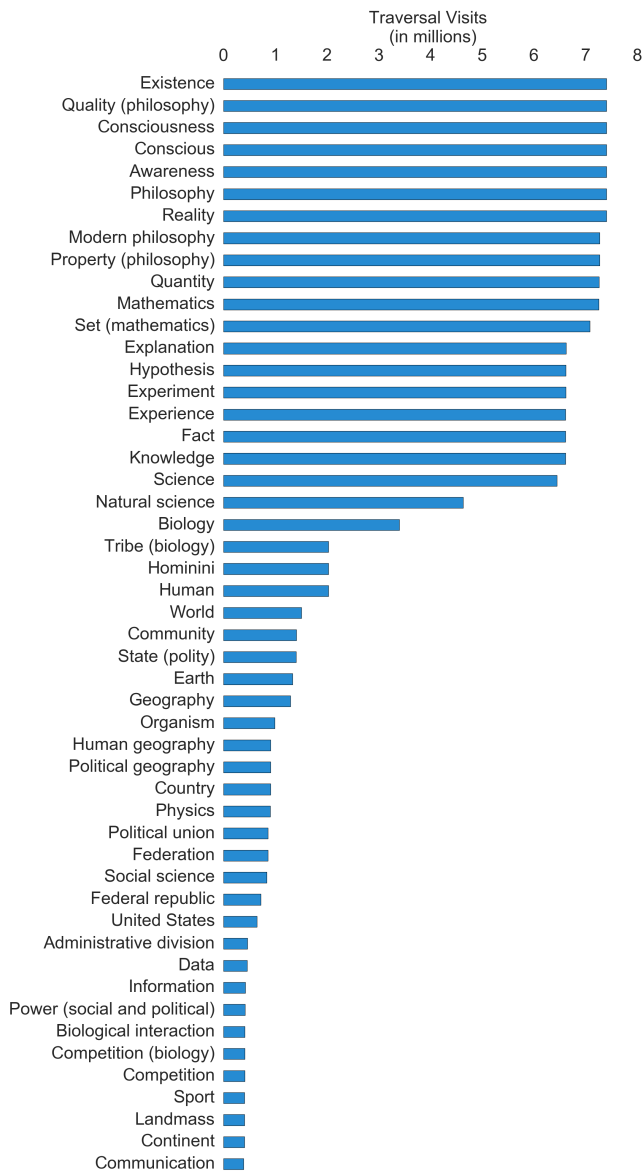}
  \caption{
    \textbf{Highest ranking articles by number of traversal visits.}
We compute the number of traversal visits for each article in the FLN (see 
Sec.~\ref{Traversal Algorithm}). 
In doing so, we can rank each article by the accumulation of first links.
The highest ranking articles by traversal visits reveal where the greatest accumulation occurs.}
  \label{fig:highest visits}
\end{figure}

\subsection{Traversal Visits}

As a distribution, the number of traversal visits by article appears to follow a decaying power law. 
The majority of articles have fewer than 30 traversal visits and
first link references accumulate at a few articles.
Specifically, $99.76\%$ of articles have fewer than $100$ traversal visits; nearly $80\%$ have none. 
Meanwhile, the highest ranking 30 articles have an extremely disproportionate number of traversal visits.

\begin{figure*}[tph!]
  \includegraphics[width=\textwidth]{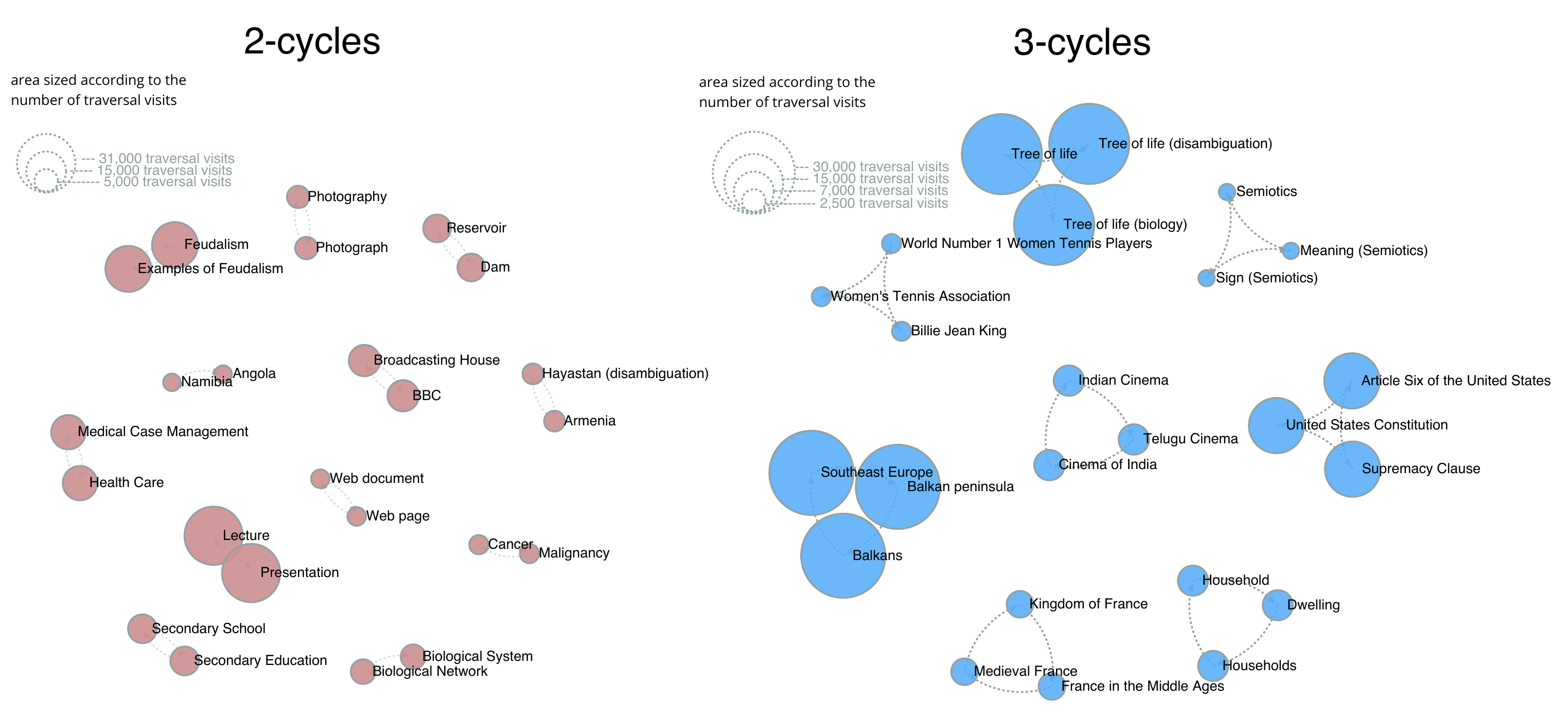}
  \caption{
    \textbf{Highest ranking 2-cycles and 3-cycles.}
We identify pairs of articles whose first links point to one another, forming
a 2-cycle. We then rank each pair of articles by the total number of 
traversal visits to gauge the most referenced groups of two articles linked
to each other. We find 2-cycles often capture synonyms or articles representing nearly the 
same concepts as opposed to distinct concepts. Similarly, we identify and 
rank 3-cycles to find they appear to capture three closely related ideas or synonyms.}
  \label{fig:cycles}
\end{figure*}
In Fig.~\ref{fig:Distribution of Visits}, we fit traversal visits against a power law distribution using the Maximum Likelihood Estimator with xmin as the number of traversal visits minimizing Kolmogorov-Smirnov distance. We find xmin to be 866 with a corresponding power law rank exponent $\alpha \simeq 1.72$ and a size exponent $\gamma \simeq 1.58$.
We also compared the fit against alternative distributions and found signficiant likelihood ratios favoring a power law fit to an exponential ($R>0, p=2.5\times10^{-60}$) and stretched exponential ($R > 0, p=.0001$).
Similarly to in-degree, we found an inconclusive comparison against a lognormal distribution and a signficiant loglikelihood ratio favoring a truncated power law given the max number of traversal visits is restricted by the number of articles in the FLN. 
On the aggregate, the distribution of traversal visits suggests
a handful of the highest ranking articles contain a disproportionate number of traversal visits, while most have none. The skew in the distribution is not terribly surprising when considering the heuristic of how the links flow: from specific to general. 

The highest ranking articles by traversal visits are broad topics spanning academic disciplines 
or notions fundamental to society~(Fig. \ref{fig:highest visits}): ``Science,'' ``Mathematics,'' ``Geography,'' and ``Philosophy''
as well as ``Community,'' ``State,'' ``Earth,'' ``Information,'' ``Power,'' and ``Communication.''
The first links flow towards broader topics, accumulating at of these foundational notions.
While ``Banana'' is a concrete fruit for example, 
the first links flow from ``Fruit'' to ``Botany'' to ``Biology,'' and ultimately 
culminate with ``Science'' and ``Philosophy.'' 
We observe that first links direct the wealth of specific knowledge on Wikipedia to 
a few foundational notions.

\subsection{Network Cycles}
\label{cycles}

We first identify 2-cycles, meaning a pair of articles with first link pointing to one another.
Of the 11 million articles, roughly 84,000 are members of 2-cycles. 
The highest ranking 2-cycles by traversal visits tend to be synonyms (or nearly so) rather than distinct, yet connected topics:
``Health Care'' and ``Medical Case Management''; ``Broadcasting House'' and ``BBC''; and ``Secondary Education'' and ``Secondary School'' 
(see Fig.~\ref{fig:cycles}).

Outside of these prominent 2-cycles, the typical 2-cycle signals a connection between distinct, yet very closely related concepts. 
For example, we observe link patterns such as inventor to product (``Voere'' to ``VEC-91''), event to organizer (``Poetry Bus Tour'' to ``Weave Books''), and book to author (``Anatomy of Britain'' to ``Anthony Sampson'').

Similarly, 3-cycles capture a synonymous or close relation among 3 articles: ``Tree of life (Biology),'' ``Tree of life (disambiguation),'' 
and ``Tree of life''; ``Cinema of India,'' ``Indian Cinema,'' and ``Telugu Cinema''
(see Fig.~\ref{fig:cycles}).
Once we extend our cycle size beyond a length of 6 however, 
``Philosophy'' along with the remaining list of high ranking articles by traversal visits dominate.
The longest cycle in the network spans 365 articles of Eastern Orthodox Liturgics for each calendar day.
Other lengthy cycles span 60--75 articles including collections of articles on national histories such as ``Japanese Eras'' 
or judicial bodies such as the ``Legislative Assembly of Ontario.''

\subsection{Path-Connected Articles}

The highest ranking groups of path-connected articles (Sec.~\ref{Traversal Algorithm}) by the number of traversal visits are articles around ``Philosophy.'' 
The highest ranking paths include branches of philosophy flowing through 
``Awareness,'' ``Existence,'' and ``Consciousness'' to ``Philosophy,'' 
which form a cycle of seven articles. As of April 10, 2016 the path-connected group around
``Philosophy'' has changed to contain instead ``Ancient Greece,'' ``Greek,'' and 
``Indo-European Languages'' (see network resilience Sec.~\ref{Traversal Funnels}).
Other paths
include concepts around ``Mathematics,'' ``Scientific,'' ``Experiments,'' 
``Biology,'' and ``Fact.''
These paths link many specific articles to ``Philosophy,'' each funneled through a particular domain.

Moving beyond ``Philosophy,'' we find other groups around 
foundational concepts such as ``Community,'' ``Landmass,'' ``Federal Government,'' 
``Presentation,'' and ``Belief System.'' 
The groups around each of these foundational notions are 
contain closely related articles. For example the group connecting
``Community''
flows from ``United States'' to ``Federal Republic'' to ``Political Union'' to ``State'' culminating at ``Community''; we also find groups flowing from 
``Public Policy'' to ``Executive (government)'' to ``Government'' to ``State'' and then 
to ``Community''; we also find paths flowing through a similar chain beginning
with ``Democracy,'' another beginning with ``Constitution,'' another at 
``Dictatorship,'' and so on. The articles build from specific means of organizing
a community (or society) and then build up to ``Community.'' 
Other path-connected groups around landmass for example begin at specific geographical regions
such as ``Eastern Europe'' building up to ``Continent'' and finally ``Landmass''.

\begin{figure}[tp!]
  \includegraphics[width=\columnwidth]{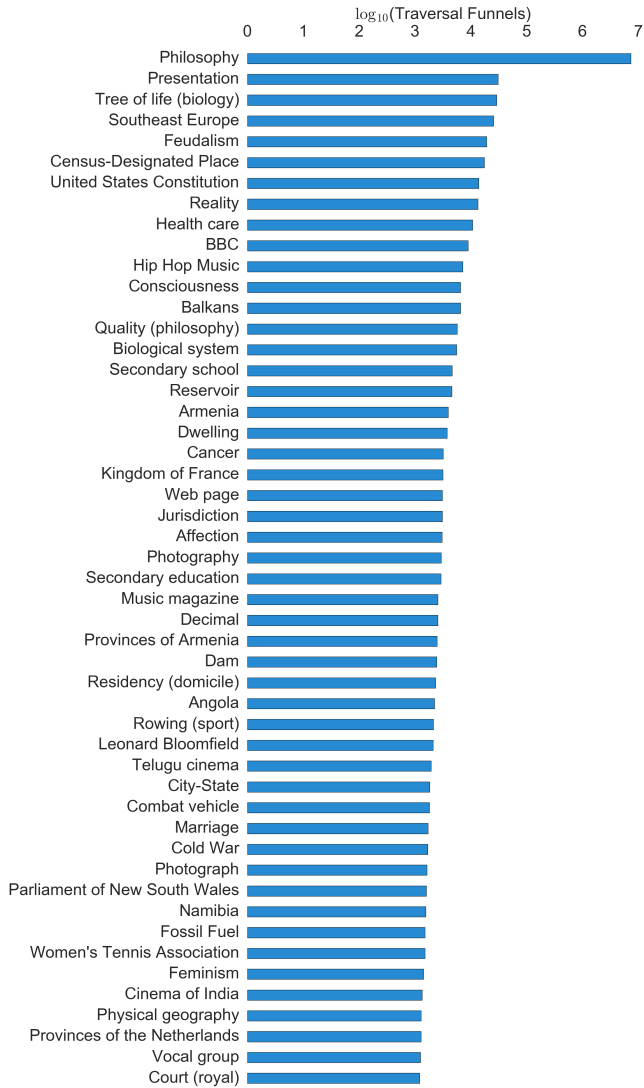}
  \caption{
    \textbf{Funnels.}
We represent the highest-ranking articles by the number of traversal 
funnels to gauge the influence each article exerts in shaping the 
structure of the FLN. 
We find ``Philosophy'' exerts an overwhelming proportion
of the influence, with other abstract notions and topical concepts ranking
next.}
  \label{fig:Funnels}
\end{figure}

\subsection{Traversal Funnels}
\label{Traversal Funnels}

\begin{figure}[tp!]
  \includegraphics[width=\columnwidth]{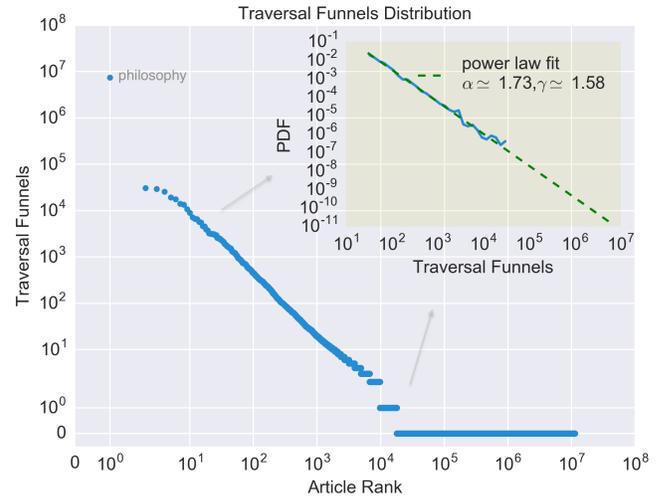}
  \caption{
    \textbf{Distribution of Traversal Funnels.}
    We plot the number of traversal funnels for each article by rank on a $\log_{10}-\log_{10}$. We then fit the data to a power law 
    distribution using the Maximum Likelihood Estimator with xmin as the number of traversal visits minimizing Kolmogorov-Smirnov distance. We find xmin to be 30 with a corresponding power law rank exponent $\alpha \simeq 1.73$ and a size exponent $\gamma \simeq 1.58$.
    The outlier on the top left is Philosophy with more traversal funnels than the second highest ranking article by two orders of magnitude.
}
  \label{fig:Funnels Distribution}
\end{figure}

To analyze the influence an article exerts in shaping the 
structure of the FLN, we compute the number of traversal funnels for each.
Ranking articles by the number of traversal funnels we find 
``Philosophy'' to be by far the highest ranking article with 
$7.37$ million paths
(see Fig.~\ref{fig:Funnels}).
Of any article, the number of traversal funnels Philosophy holds exceeds 
all others by more than two orders of magnitude.
The ``Philosophy'' cycle which contains ``Existence,'' ``Awareness,'' ``Reality,'' 
and similar articles accumulates the overwhelming proportion of its 
references through ``Philosophy'': $7.37$ million of the $7.4$ million references
are funneled through ``Philosophy''.
Second on the list of highest-ranking articles by traversal funnels is 
``Presentation'' with only $30$ thousand paths. Similarly abstract 
concepts also rank highly such as ``Tree of life'' (30 thousand), 
``Reality'' (13 thousand), and ``Jurisdiction'' (3 thousand).

Many high-ranking articles are remarkably topical, culturally and politically important concepts.  For example, ``Health Care,'' a recently high-contested legislative topic appears high on the list---Google trends indicates an uncharacteristic spike in search frequency between August, 2009 and February, 2010.
Other high ranking articles include key historical events such as the ``Cold War'' or critical scarce resource with recent 
media discussion such as ``Fossil Fuel.'' 
The highest-ranking list also includes ``Hip Hop,'' ``Cancer,'' and ``Web Page.''

In contrast to traversal visits, traversal funnels are comparatively resilient to changes in the FLN over time. 
While a few links can alter a cycle, potentially displacing an article that was the recipient of references, traversal funnels remain stable even when cycles change.

As a distribution, we find few articles influence the structure of the 
FLN. Only $17, 821$ articles have one or more traversal funnels, leaving
more than $99\%$ with none---most articles are recipients of 
the references flowing through the articles with at least one traversal funnel.
When to a power law 
    distribution (using the Maximum Likelihood Estimator) with xmin as the number of traversal funnels minimizing Kolmogorov-Smirnov distance), we find xmin to be 30 with a corresponding power law rank exponent $\alpha \simeq 1.73$ and a size exponent $\gamma \simeq 1.58$
(see Fig.~\ref{fig:Funnels Distribution}). 
When compared against alternative distributions, we find inconclusive results for lognormal and truncated power law distributions and significant log likelihood ratios favoring a power law for exponential ($R>0, p=5.3\times10^{-6}$) and stretched exponential ($R>0, p = 1.8\times10^{-8}$) distributions.
Even within the few articles
influencing the structure of the FLN, only a handful of these exert most of the 
control. 

\subsection{Article Popularity}

\begin{figure}[tp!]
  \includegraphics[width=\columnwidth]{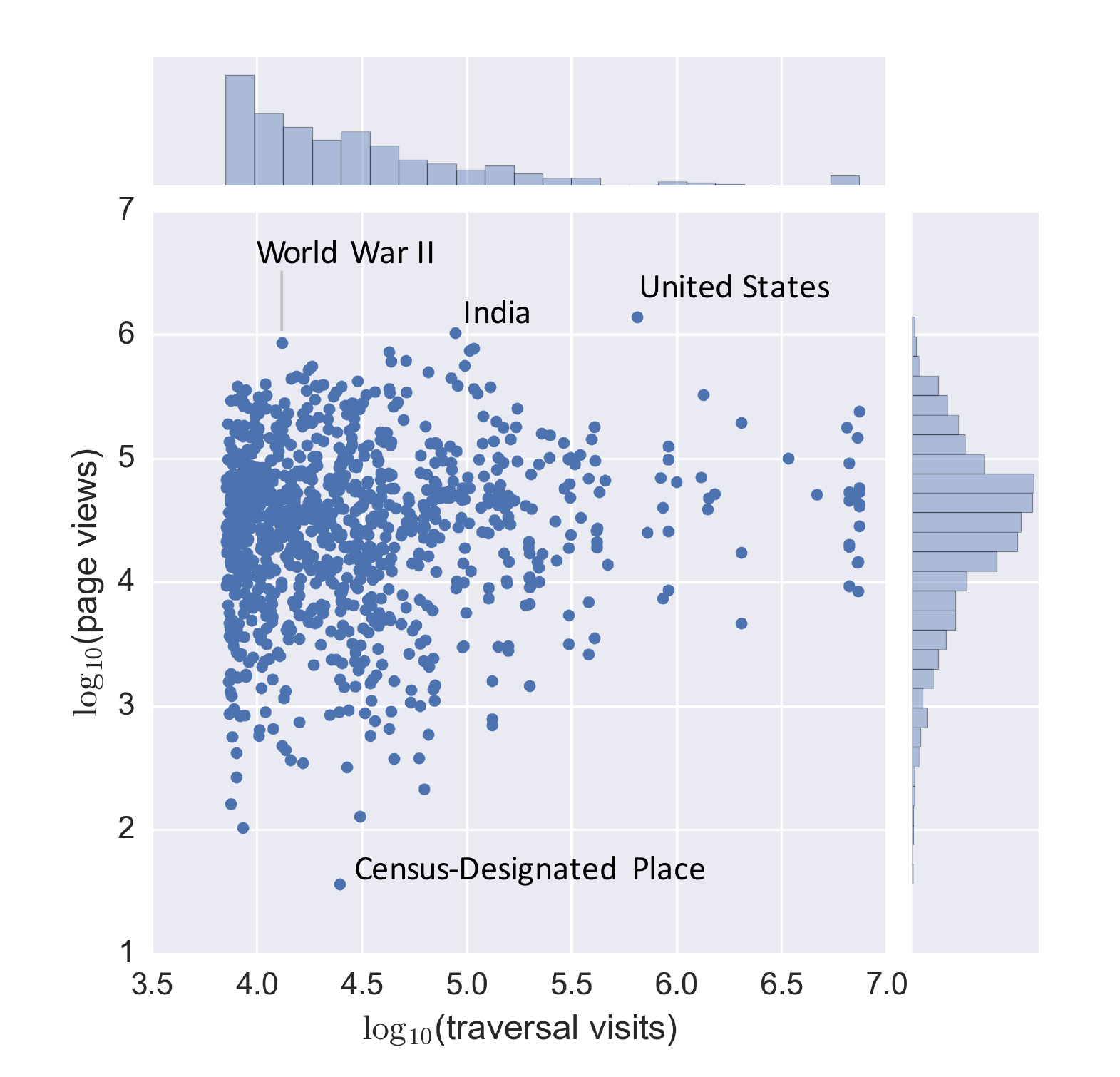}
  \caption{
\textbf{Article Popularity (by page views) for the highest ranking 1000 articles by traversal visits.}
We use the total number of page views provided by Wikipedia for the month
of October 2015 to compare each article's popularity against traversal visits.
Overall, more page views do not correspond to higher traversal visits---Wikipedia users do not tend to visit articles with a higher accumulation in the FLN more so than they visit others. 
However, the variation in popularity appears to decrease as the number of traversal visits increases. The greatest number of articles fall within roughly 1 million traversal visits and page views. Furthermore, page views for the top 1000 articles articles by traversal visits appear to be log-normally distributed. 
}
  \label{fig:Views and Visits}

\end{figure}
\begin{figure}[tp!]
  \includegraphics[width=\columnwidth]{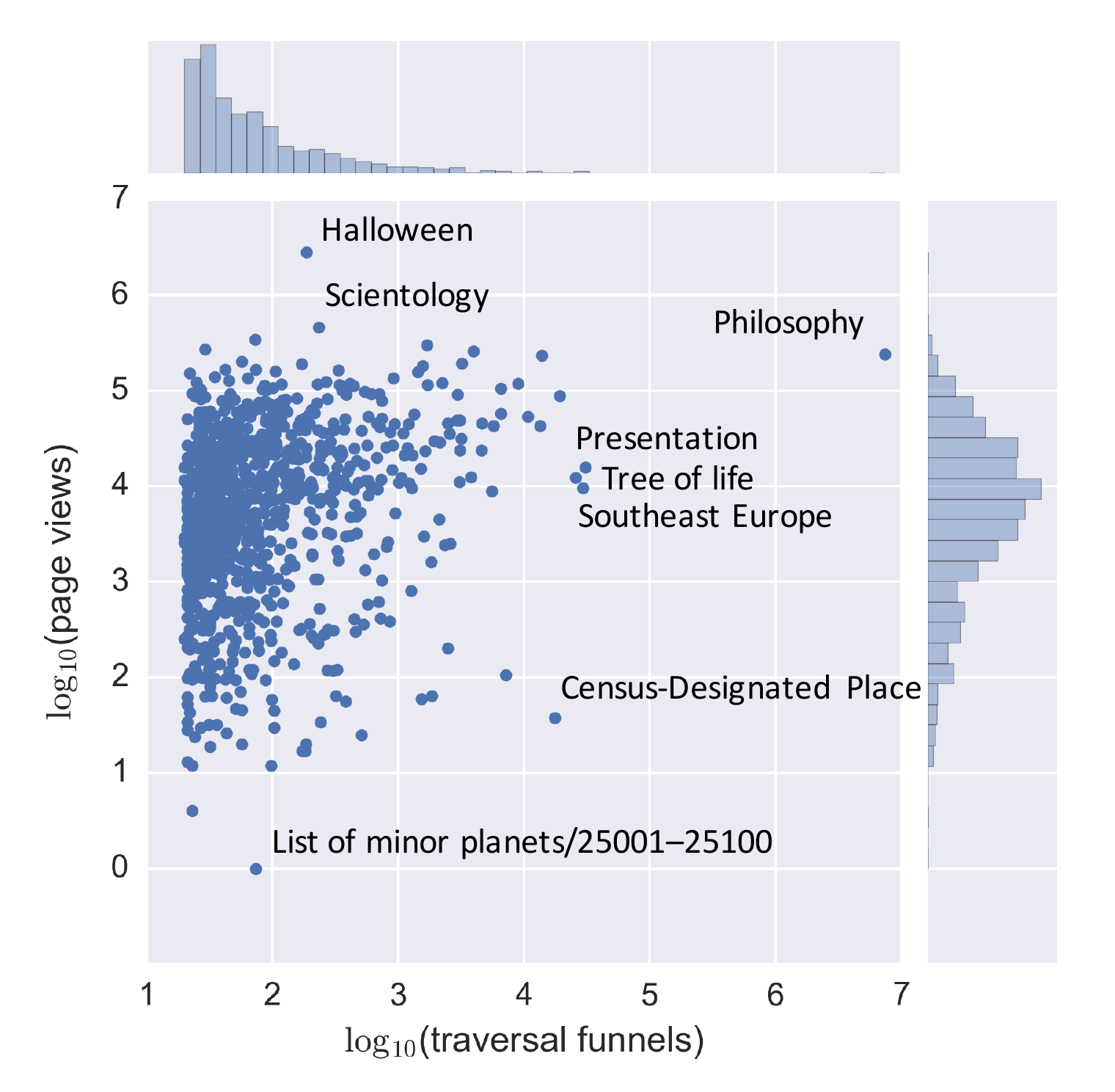}
  \caption{
\textbf{Article Popularity (by page views) for the highest ranking 1000 articles by traversal funnels.}
We use the total number of page views provided by Wikipedia for the month
of October 2015 to compare each article's popularity to traversal funnels.
Similar to traversal visits, more page views do not correspond to higher traversal funnels---Wikipedia users do not tend to visit articles with a higher influence in the FLN more so than they visit others.
}
  \label{fig:Views and Funnels}
\end{figure}

Wikipedia released an API to measure article popularity by page views
starting November 2015. 
Page views add another dimension to our
findings by constrasting
the number of users who access a particular article against
the number of traversal visits and funnels.
We measure popularity as the total number 
of page views in the English edition of Wikipedia in the month of 
October 2015---the earliest full month for which the data is available. 
We find the highest ranking 1000 articles (by traversal visits) have an average of
$70$ thousand page views in October with high variation: the standard deviation 
is $1.1$ million page views. 
The number of page views has a skewed distribution with $75\%$ of articles
reaching fewer than $73$ thousand page views in October.
The article for ``United States'' has the most page 
views of the highest ranking $1000$ articles by traversal visits with 
$1.4$ million views in October. The next most popular articles are 
``India,'' ``World War II,'' and the ``United Kingdom'' each with roughly 1 million page views. 
The ``Philosophy'' article, despite outranking every article by traversal visits,
has only $240$ thousand page views in October.

We also analyze article popularity for the highest ranking 1000 articles by 
traversal funnels. In October, the average page views per article is 
$22$ thousand with a standard deviation of $95$ thousand views. The distribution
is skewed with $75\%$ of articles reaching fewer than $20$ thousand page views. 
``Halloween'' is the most popular article with $2.8$ million views in October,
likely a result of Halloween falling in the month of October. 
The second most popular article, ``Scientology'' ($463$ thousand views) appeared in popular news as actress Leah Remini announced she's leaving the Church of Scientology 
\cite{scientology}.
Other popular articles include ``Clint Eastwood'' ($341$ thousand views),
and the ``Cold War'' ($298$ thousand views) although each has significantly fewer views compared to the views for ``Halloween''.
Other standout popular October articles include ``24-hour'' and ``12-hour'' clocks likely due
to Daylight savings in October and ``Marriage'' possibly due to the drop in the number of
weddings in the months following October
\cite{weddings}.
``Philosophy'' which ranks seventh among the most popular of the highest ranking articles by traversal funnels, appears nowhere in the top 20
by traversal visits with only $240$ thousand page views in October, 2015. ``Philosophy'' is a relatively popular article 
among articles influencing the shape of the network, but less popular among 
highly-ranked articles by accumulation.

Overall, the contrast between page views and traversal visits or funnels highlights a difference between popularity and influence in the FLN. Although traversal funnels and visits point to influential articles where many topics culminate, Wikipedia users do not visit these articles more than others. What is popular, is not necessarily foundational, and vice versa.


\section{Concluding Remarks}

The findings here should only be considered within the limitations of their context.
We examined only the English edition of Wikipedia at a particular moment in time, 
considering only the first link in in each article as a means to relate articles.
Finally, Wikipedia, while the largest digital 
collection of human knowledge, is rife with the biases of the many contributing editors
\cite{wagner2015s}.
Nevertheless, the findings do reveal
generalizable relationships and point to foundational notions.

Among our observations is the appearance of multiple scale-free distributions within the network. 
Few articles have most traversal visits, few paths have an exceptionally long path length, and even fewer
articles are responsible for funneling most paths. When measured against the traversal funnels, 
``Philosophy'' emerges as an exceptional article by orders of magnitude. 
Nevertheless, many other foundational concepts emerged naturally within FLN. 
Path connected-groups around ``Community'', ``State'', and ``Science'' reveal a foundational structure within the network. 
More curious is the emergence of recently prominent political and economics topics such as ``Fossil Fuel'' and ``Health Care'' 
within the highest ranking funnels. 
Wikipedia seems to reflect not only timeless foundations, but also the topical (at least within English speaking society).

Future work could examine other language versions of Wikipedia for potentially telling cultural or regional differences as well as expand the network to more than the first link.
In addition, the work provides a general algorithm for assessing the structure and measuring influence in a directed network (or any braided flow structure). 
Traversal funnels is a new measure of influence for directed networks without spill-over into cycles in contrast to traditional network centrality measures.
When applied to Wikipedia, these findings also form the basis for the creation of a taxonomy where 
every idea, event, or object sits within a hierarchy of connected notions.
The taxonomy would extend a traditional word thesaurus beyond mere synonyms to a related hierarchy of concepts.
Applications could range from an enhanced network of ideas to psychological insights into how humans form associations.
Specifically, an ever-evolving reference of related hierarchical concepts could be 
used to improve search engine algorithms or natural language processing.

\newpage

\section{Appendix}
\label{Appendix}

\subsection{Constructing The First Link Network}

To map Wikipedia's First Link Network, we use the freely-available XML dump of the English edition of Wikipedia. 
Rather than rely on a sample of articles from which to generalize, we opted to process the entirety of Wikipedia, 
eliminating any statistical error due to sampling.
We analyze the snapshot provided on November 2014, representing the state of Wikipedia at the time.
The November raw dump consists of 11 million articles: 4.7 million unique articles along with redirects
and disambiguations.
Knowing Wikipedia is an ever-evolving project with 10 edits every second and 750 new articles per day on average
\cite{wiki_edits},
our aim is to characterize the structure of the First Link Network.

Wikipedia renders and stores articles in MediaWiki markup, a markup language with syntax and keywords to format and mark elements in a page. Along with special syntax for links, MediaWiki markup includes templates for audio files, images, and side-bar
information.
While a human could manually identify the first link, to map the entire First Link network of 11 million articles, we needed to programmatically untangled the body text from side-bar, header box, and invalid link elements.

While some libraries exist for MediaWiki Markup,
approaches using existing libraries led to several bugs 
including trouble with nested links, nested parenthesis, unclosed tags, escape characters 
as well as compatibility with other libraries used to parse the XML.
Consequently,  we developed an algorithm for parsing the first link in the XML version of each article.
Our parsing algorithm aimed to: 
1) accurately identify the first link among other page elements, and 
2) efficiently do so---that is without needing for several passes through the data.
To process an article in a single pass, we developed a hierarchical system of flags:
\begin{figure}[tp!]
  \includegraphics[width=\columnwidth]{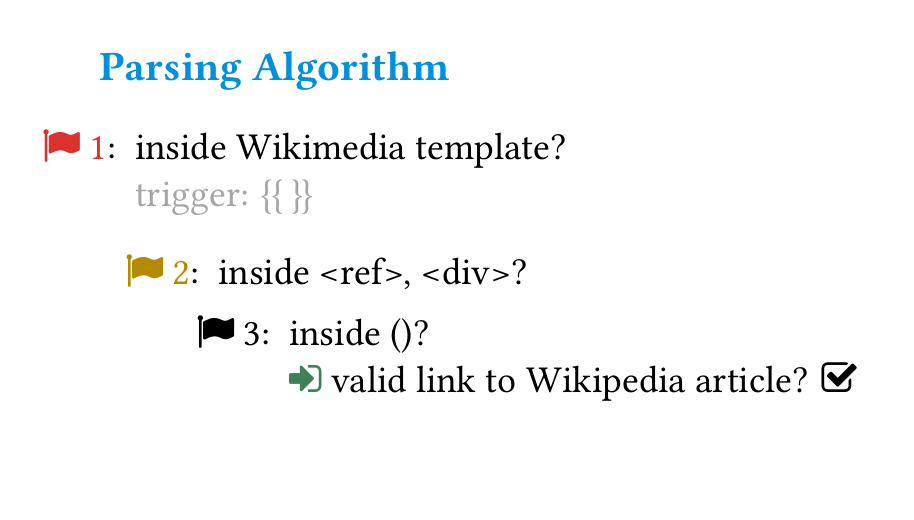}  
  \caption{
    \textbf{Parsing Algorithm of Wikipedia's XML dump.}
The highest flag in the hierarchy indicates a Wikimedia template used to mark an element in the side bar, display an image, link to an external file, or another Wikimedia project outside of Wikipeida. Next, to catch any remaining elements outside the main body we have a second flag for <ref>, <div> elements. Finally, we identify parenthesis to ensure we do not capture a link to a pronunciation key.}
  \label{fig:parsing algorithm}
\end{figure}

The algorithm loops in three-character chunks to account for potentially nested elements, 
shifting by one character steps through the article markup.
If any markup triggers for a flag are detected, a flag is raised. 
Once a flag is raised, we stop processing and proceed to the next character
until the flag's closing markup.
A first link is identified only if Flags 1, 2, and 3 are all off.
In this case, the entire link is retrieved. 
We then confirm the link is valid by filtering for MediaWiki keywords indicating external page or other projects
as well as common file extensions for 
images, audio files, and the like 
\cite{media_wiki_templates}.
The first link of an article is then the earliest valid link with unraised flags.

To process the entirety of Wikipedia, we distributed the parsing and processing of the XML dump
across 112 cores of the UVM supercomputer cluster
\cite{vacc}.
We then joined the results to form a hash table containing every Wikipedia article and its corresponding
first link. The resulting network map is the basis of our analysis.
An online appendix containing all of the code and data used can be accessed \href{http://compstorylab.org/share/papers/ibrahim2016a/index.html}{here}.

\subsection{Power Law Alternative Distributions}
\label{power_law_comp}

Using the powerlaw Python package developed by Alstott, we apply MLE to fit a power law exponent and compare the fit to alternative distributions
\cite{clauset2009power, alstott2014powerlaw}.
For in-degree, we find a significant log likelihood ratio favoring a  truncated power law ($R < 0, p=4.5\times10^{-13}$) given the maximum in-degree is limited by the number of articles in the FLN. We find an inconclusive log likelihood ratio for a lognormal distribution and significant log likelihood ratios favoring a power law fit against an exponential ($R  > 0, p=0$) and against a stretched exponential ($R > 0, p = 3.0\times10^{-50}$). 

\acknowledgments
MI is grateful to RJ for pointing out the Reddit post 
\cite{reddit}
describing how the majority of links lead to ``Philosophy,'' inspiring this research.

The authors are also grateful for the suggestions and input provided by Randall Harp.

PSD and CMD acknowledge support from NSF Big Data Grant $\#1447634$.

The authors also acknowledge support from the Vermont Advanced Computing Core which is supported by NASA (NNX 06AC88G), at the University of Vermont for providing High Performance Computing resources that have contributed to the research results reported within this paper.

\clearpage

\end{document}